\documentclass[%
 aip,
 amsmath,amssymb,
preprint,%
]{revtex4-2}

\usepackage{graphicx}% Include figure files
\usepackage{dcolumn}% Align table columns on decimal point
\usepackage{bm}% bold math
\usepackage[utf8]{inputenc}
\usepackage[T1]{fontenc}

\usepackage{hyperref}
\usepackage{braket} % Introduced by the author
\usepackage{comment} % Introduced by the author
\usepackage{colortbl}
\definecolor{Gray}{gray}{0.9}

\newcommand{\be}{\begin{equation}}
\newcommand{\ee}{\end{equation}}
\newcommand{\ti}[1]{\text{#1}}
\newcommand{\mc}[1]{\mathcal{#1}}
\newcommand{\hmc}[1]{\hat{\mathcal{#1}}}

\newcommand{\eq}[1]{Eq.~\eqref{#1}}
\newcommand{\eqs}[2]{Eqs.~\eqref{#1} and \eqref{#2}}
\newcommand{\Eq}[1]{Equation~\eqref{#1}}

\newcommand{\stn}[1]{Sec.~\ref{#1}}
\newcommand{\Stn}[1]{Section~\ref{#1}}
\newcommand{\app}[1]{Appendix~\ref{#1}}

\newcommand{\bea}{\begin{eqnarray}}
\newcommand{\eea}{\end{eqnarray}}
 
\newcommand{\ba}{\begin{array}}
\newcommand{\ea}{\end{array}}

\newcommand{\bl}{\begin{flalign}}
\newcommand{\enl}{\end{flalign}}

\newcommand{\tr}{\text{Tr}}

\usepackage[normalem]{ulem} % For strikeout

\newcommand{\mkerase}[1]{\ifmmode{\color{red}{\text{\sout{\ensuremath{#1}}}}}\else{\color{red}{\sout{#1}}}\fi}

\begin{document}

\preprint{AIP/123-QED}

\title{General Framework for Quantifying Dissipation Pathways in Open Quantum Systems. I. Theoretical Formulation}

\author{Chang Woo Kim}
\affiliation{
    Department of Chemistry, Chonnam National University, Gwangju 61186, South Korea
    }
\email{cwkim66@jnu.ac.kr}
    
\author{Ignacio Franco}
\affiliation{
    Department of Chemistry, University of Rochester, Rochester, New York 14627, USA
    }
\affiliation{
    Department of Physics, University of Rochester, Rochester, New York 14627, USA
    }
\email{ignacio.franco@rochester.edu}

\date{\today}% It is always \today, today,
             %  but any date may be explicitly specified

\begin{abstract}
We present a general and practical theoretical framework to investigate \emph{how} energy is dissipated in open quantum system dynamics. This is done by quantifying the contributions of individual bath components to the overall dissipation of the system. The framework is based on the Nakajima-Zwanzig projection operator technique which allows us to express the rate of energy dissipation into a specific bath degree of freedom by using traces of operator products. The approach captures system-bath interactions to all orders, but is based on second-order perturbation theory on the off-diagonal subsystem's couplings and a Markovian description of the bath. The usefulness of our theory is demonstrated by applying it to various models of open quantum systems involving harmonic oscillator or spin baths, and connecting the outcomes to existing results such as our previously reported formula derived for locally coupled harmonic bath [J.~Chem.~Phys.~\textbf{154},~084109~(2021)]. We also prove that the dissipation calculated by our theory rigorously satisfies thermodynamic principles such as energy conservation and detailed balance. Overall, the strategy can be used to develop the theory and simulation of dissipation pathways to interpret and engineer the dynamics of open quantum systems.
\end{abstract}

\maketitle

\section{Introduction}\label{section:introduction}
A realistic dynamical process always involves exchange of energy between the central degrees of freedom (subsystem) and the surrounding environment (bath), which is referred to as dissipation.\cite{Weiss2012,Schlosshauer2007} This is particularly important in the condensed phase, in which thermal fluctuations of the bath often profoundly affect the dynamics of the subsystem. For example, quantum transport\cite{Fassioli2014,Wang2015,Hestand2018,Mejia2022} of electrons or molecular excitations along extended molecules or molecular arrays often occur via energetic relaxation of the subsystem, with the excess energy dissipated to satisfy the energy conservation. For this reason, the total dissipated energy reflects the progress of the relaxation during the dynamics. In particular, it is highly desirable to understand the relative importance of each bath degree of freedom (DOF) at each instance or the overall relaxation process. This can be accomplished by decomposing the dynamics of the total dissipated energy into individual contributions by specific components of the bath. Such an ability will provide detailed knowledge about which bath components play a predominant role in the dynamics, which can help us to establish useful principles for designing and controlling the quantum behavior of molecular systems.\cite{Kienzler2014,CamposGonzalezAngulo2019,Ng2020,Hart2021} For instance, such insights are needed to understand how energy is dissipated by the photosynthetic complexes, and how to modify physical systems to enhance or suppress dissipation.

In the field of quantum thermodynamics, methodologies for analyzing the flow of heat and entropy between macroscopic bath reservoirs have been developed for closed\cite{Esposito2006,Esposito2010} and driven systems.\cite{Kato2016,Kosloff2019,Carrega2016} These tools can provide detailed descriptions about how thermodynamic principles manifest in quantum systems such as quantum heat engines\cite{Geva1996} and rectifiers.\cite{Mascarenhas2016} However, there are still relatively few methods which can efficiently track time dependence of the quantum statistical properties related to individual microscopic components of the bath. In this paper, we address this challenge by extending the theory of quantum master equations, which is one of the major workhorses for simulating the dynamics of open quantum systems.

To calculate the dissipation due to a single bath component, we need to monitor its change in energy which, in turn, requires dynamical information about the environment. In principle, this task can be fulfilled by utilizing already available simulation methods for quantum dynamics that follow the bath explicitly. In practice, this remains a widely challenging task. For example, the energy of a single DOF can be straightforwardly obtained from efficient wavefunction propagation methods such as multi-configurational time-dependent Hartree (MCTDH)\cite{Meyer1990,Worth2000,Wang2015a} or time-dependent density matrix renormalization group (TD-DMRG).\cite{Cazalilla2002,White2004,Haegeman2016} However, incorporating the effect of temperature requires additional complications such as combination with thermofield dynamics.\cite{Umezawa1982} Even so, the bath subspace must be truncated to reduce the computational burden to an amenable extent,\cite{Ren2018,Jiang2020} which can significantly affect the accuracy of the calculation.

Another major category of simulation methods are quantum master equations where the focus is on the reduced density matrix (RDM) of the subsystem and the influence of the environment is only captured implicitly by determining how it affects the subsystem. For these methods, the dynamics of a bath component can be accessed by combining the mode with the subsystem and propagating the RDM of this extended subsystem.\cite{O'Reilly2014,Nalbach2015,Novoderezhkin2017,Bennett2018} Even when the simulation method does not easily allow such a re-definition of the subsystem, the dissipation can be indirectly extracted by introducing an additional bath mode into the subsystem, which acts as a probe.\cite{Kim2022} Nevertheless, both procedures can be computationally quite demanding, as the construction of the extended subsystem significantly increases the dimension of the RDM. This issue becomes especially problematic when the bath frequency becomes comparable to or less than the thermal energy, where a large number of bath quantum states must be explicitly included in the density matrix to faithfully simulate the statistical mechanics of the bath mode.

Mixed quantum-classical simulation methods,\cite{Kim2008,Kelly2015,Liu2018,Runeson2020} where the bath is captured classically, can be used to compute the energy of the bath modes with only modest computational costs.\cite{Kim2020,Cho2021} However, the reliability of the calculation is often deteriorated by the approximations involved in the formulation. Indeed, these methods often exhibit uncontrollable artifacts such as negative subsystem state populations or zero-point leak of the vibrational energy.\cite{Kim2020,Kim2014}

Motivated by the absence of efficient yet reliable tools for resolving dissipation, we have recently developed a practical scheme\cite{Kim2021} based on Fermi's golden rule rate in the weak-coupling limit. The scope of our theory covers non-adiabatic chemical dynamics in the condensed phase such as the transfer of molecular excitation\cite{Forster1959,Sumi1999} or charge,\cite{Song1993,Evans1996} in the presence of harmonic bath modes affecting the process. Although the master equation for the population of the subsystem states was already reported several decades ago,\cite{Mukamel1983,Sumi1999} our previous work extended it to resolve the overall dissipated energy into the amounts absorbed by each bath mode. This was accomplished by explicitly quantizing a bath mode and calculating how its vibrational energy changes upon the population transfer between electronic-vibrational (vibronic) quantum states. Deriving an efficient expression, however, required a clever but not readily generalizable analytical summation over all vibronic state pairs based on an equality extracted from the theory of spectral line shapes.\cite{Hayes1988}

In this paper, we present a general framework for constructing practical and accurate schemes to isolate dissipation pathways in open quantum systems. The formulation incorporates a specific bath DOF into the subsystem component and calculates the change in its energy with Nakajima-Zwanzig projection operator technique.\cite{Nakajima1958,Zwanzig1960} By perturbatively expanding the Liouville-von Neumann equation, we elucidate the rate of dissipation expressed by using traces of operator products, which can be applied for general types of bath and subsystem-bath interaction. We also rigorously prove that this approach satisfies energy conservation and detailed balance. We then demonstrate the applicability of our approach by using it toward deriving the dissipation rate equations for prototypical models of open quantum systems, namely subsystems coupled to harmonic oscillator baths or spin baths. In the subsequent paper, Paper II,\cite{Kim2024} we will use the developed expressions to quantify the dissipation pathways in model Hamiltonians and assess the accuracy of the theory by benchmarking it against numerically exact simulations.

%Additionally, in this paper, we assess the reliability of our theory by comparing it to numerically exact results that were unavailable in our earlier work. Finally, we demonstrate that incorporating non-Markovianity significantly enhances the accuracy of our method and enables us to achieve nearly quantitative decompositions of the dissipated energy.

The structure of the paper is as follows: In \stn{section:background}, we first provide an overview of the theoretical background required to understand the main findings of our work, and introduce our new framework for quantifying the dissipation pathways. In \stn{section:specific}, we apply our framework to specific model Hamiltonians and connect the outcomes to previously known results. In \stn{section:Conclusion} we summarize our main findings and suggest future research directions.

%\Stn{section:accuracy} evaluates the precision of the derived formulas and investigates the factors influencing the frequency-dependent dissipation of energy. We also show that introducing non-Markovianity enhances the reliability of our developed method. Finally, in \stn{section:Conclusion}, we conclude the paper by summarizing our discoveries and proposing potential avenues for future research.

\section{Theory}\label{section:background}
Our objective is to extract the rate of dissipation into a specific bath DOF, under the dynamics governed by a quantum master equation. As our approach is developed based on the projection operator technique, we present its brief review in \stn{subsection:Hamiltonian}. We then propose our new theoretical framework for resolving the dissipated energy into individual bath components in \stn{subsection:calc_diss}. Finally, \stn{subsection:proof} discusses energy conservation and detailed balance.

\subsection{The projection operator technique for open quantum systems}\label{subsection:Hamiltonian}
We consider a group of quantum states interacting with the surroundings, which are classified as the subsystem and bath, respectively. We adopt the viewpoint of open quantum system and divide the Hamiltonian $\hat{H}$ of the system as
\begin{equation}\label{eq:H_global}
    \hat{H} = \hat{H}_\ti{sub} + \hat{H}_\ti{bath} + \hat{H}_\ti{int},
\end{equation}
where $\hat{H}_\ti{sub}$ is the Hamiltonian of the subsystem, $\hat{H}_\ti{bath}$ the bath, and $\hat{H}_\ti{int}$ the interaction between the subsystem and bath.

We cast $\hat{H}_\ti{sub}$  by using $\{\ket{A}\}$ as the basis and split it into diagonal and off-diagonal components which account for the state energies and inter-state couplings, respectively:
\begin{equation}\label{eq:H_sys}
    \hat{H}_\ti{sub} = \hat{H}_\ti{ener} + \hat{H}_\ti{coup},
\end{equation}
\begin{equation}
    \hat{H}_\ti{ener} = \sum_A E_A \ket{A}\bra{A},
\end{equation}
\begin{equation}
    \hat{H}_\ti{coup} = \sum_{A} \sum_{B<A} V_{AB} \ket{A} \bra{B} + \ti{H.c.}
\end{equation}
Here, $E_A$ is the energy of the state $A$, $V_{AB} = V_{BA}$ is the coupling between states $A$ and $B$, and H.c. denotes Hermitian conjugate. In turn, we assume that the sum of the Hamiltonian components for bath and subsystem-bath interaction can be split into individual elements $\{\hat{h}_j\}$,
\begin{equation}\label{eq:bath_component}
    \hat{H}_\ti{bath} + \hat{H}_\ti{int} = \sum_j \hat{h}_j,
\end{equation}
and each element only couples to the diagonal part of the subsystem Hamiltonian
\begin{equation}
    \hat{h}_j = \sum_{A} (\ket{A} \bra{A} \otimes \hat{v}_{Aj}).
\end{equation}
On the other hand, the off-diagonal component of $\hat{H}_\ti{sub}$ does not interact with the bath in our model, which is called Condon approximation\cite{Condon1928} in molecular systems. Such an assumption is frequently employed to construct quantum master equations for chemical dynamics in condensed phases, starting from harmonic oscillator bath\cite{Evans1996,Lai2022} to more general bath models.\cite{Golosov2001,Jang2020b}

We now apply the projection operator technique\cite{Nakajima1958,Zwanzig1960} to derive the master equation that governs the time evolution of an open quantum system. To apply the technique, one first splits the identity super-operator into $\hmc{I} = \hmc{P} + \hmc{Q}$ where $\hmc{P}$ and $\hmc{Q}$ project the full density matrix $\hat{\rho}$ onto the dynamically relevant part $\hat{\mc{P}}\hat{\rho}$ and the rest $\hat{\mc{Q}}\hat{\rho}$, respectively. As $\hmc{P}$ performs a projection, it should satisfy $\hmc{P}^2 = \hmc{P}$ and also $\hmc{P}\hmc{Q} = \hmc{Q}\hmc{P} = 0$.

The time evolution of $\hat{\rho}$ is governed by the Liouville-von Neumann equation $d \hat{\rho}(t) / dt = -i \hmc{L} \hat{\rho}(t) / \hbar$, where the Liouvillian super-operator is defined as $\hmc{L} \hat{\rho} = [\hat{H}, \hat{\rho}]$. By assuming that the density of the system is initially confined in the dynamically relevant part so that $\hmc{P} \hat{\rho}(0) = \hat{\rho}(0)$ and $\hmc{Q} \hat{\rho}(0)=0$, one can derive a formally exact expression for the dynamics of $\hmc{P} \hat{\rho}(t)$,\cite{Yang2002,Mulvihill2021}
\begin{equation}\label{eq:exact_solution}
    \frac{d}{dt} [\hmc{P} \hat{\rho}(t)] = -\frac{i}{\hbar} \hmc{P} \hmc{L} \hmc{P} \hat{\rho}(t) -\frac{1}{\hbar^2} \int_0^t \hmc{P} \hmc{L} \hmc{Q} \exp \bigg[ -\frac{i(t - t')}{\hbar} \hmc{Q} \hmc{L} \bigg] \hmc{Q} \hmc{L} \hmc{P} \hat{\rho}(t') \: dt'.
\end{equation}
We now apply perturbation theory by dividing the total Hamiltonian as $\hat{H} = \hat{H}_0 + \hat{H}_1$ where
\begin{equation}\label{eq:H_split}
    \begin{gathered}
    \hat{H}_0 = \hat{H}_\ti{ener} + \hat{H}_\ti{bath} + \hat{H}_\ti{int}, \\
    \hat{H}_1 = \hat{H}_\ti{coup},
    \end{gathered}
\end{equation}
and treating $\hat{H}_1$ as perturbation. The Liouvillian is also accordingly split as $\hmc{L} = \hmc{L}_0 + \hmc{L}_1$, where
\begin{equation}\label{eq:L_split}
    \hmc{L}_0 \hat{\rho} = [\hat{H}_0,\hat{\rho}] \ti{ and } \hmc{L}_1 \hat{\rho} = [\hat{H}_1,\hat{\rho}].
\end{equation}
We now specify the form of $\hmc{P}$ as
\begin{equation}\label{eq:P_def}
    \hmc{P} \hat{\rho} = \sum_A (P_A \ket{A} \bra{A} \otimes \hat{R}_A),
\end{equation}
where $P_A = \tr_\ti{b} \bra{A} \hat{\rho} \ket{A}$ is the population of the subsystem state $\ket{A}$ and $\tr_\ti{b}$ indicates the trace\cite{Note001} over the bath subspace. We also define $\hat{R}_A$ as the equilibrium bath density for $\hat{h}_A$,
\begin{equation}\label{eq:R_def}
    \hat{R}_A = \frac{\exp(-\beta \hat{h}_A)}{\tr_\ti{b} [\exp(-\beta \hat{h}_A)]},
\end{equation}
where $\beta = 1/k_\ti{B} T$ is the inverse temperature , and $\hat{h}_A$ is the projection of the total system Hamiltonian onto a subsystem state
\begin{equation}\label{eq:h_proj}
    \hat{h}_A = \bra{A} \hat{H} \ket{A} = E_A + \sum_j \hat{v}_{Aj}.
\end{equation}
According to \eqs{eq:P_def}{eq:R_def}, $\hmc{P}$ instantly relaxes the bath density to reference densities $\{ \hat{R}_A \}$ and also quenches any coherence between the subsystem states.

From now on, we will focus on the dynamics of the projected density matrix $\hmc{P} \hat{\rho}(t)$ [\eq{eq:P_def}] and derive the rate equation that governs the subsystem state populations. We do so by following a procedure similar to that in Ref.~\citenum{Yang2002}. Specifically, we first expand \eq{eq:exact_solution} by applying time-dependent perturbation theory and keeping the terms up to the second order in $\hmc{L}_1$. From \eqs{eq:L_split}{eq:P_def} it follows that $\hmc{P} \hmc{L}_0 \hat{\rho} = \hmc{L}_0 \hmc{P} \hat{\rho} = \hmc{P} \hmc{L}_1 \hmc{P} \hat{\rho} = 0$. If we replace $\hmc{Q}$ by $\hmc{I} - \hmc{P}$ and invoke these identities, the expression simplifies into
\begin{equation}\label{eq:2nd_order}
    \frac{d}{dt} \big[ \hmc{P} \hat{\rho}(t) \big] = -\frac{1}{\hbar^2} \int_0^t \hmc{P} \hmc{L}_1 \exp \bigg[ -\frac{i(t - t')}{\hbar} \hmc{L}_0 \bigg] \hmc{L}_1 \hmc{P} \hat{\rho} (t') \: dt'.
\end{equation}
Applying the Markov approximation and calculating $\tr_\ti{b} [\bra{A} \frac{d}{dt} \{ \hmc{P} \hat{\rho}(t) \} \ket{A} ] $ lead us to the rates of change in the electronic populations
\begin{equation}\label{eq:rate_prim}
    \dot{P}_A(t) = -\frac{1}{\hbar^2} \tr_\ti{b} \bigg[ \int_0^\infty \bra{A} \hmc{P} \hmc{L}_1 \exp( -it' \hmc{L}_0 / \hbar ) \hmc{L}_1 \hmc{P} \hat{\rho}(t) \ket{A} \: dt' \bigg].
\end{equation}
We expand the integrand in \eq{eq:rate_prim} with the expressions for the Liouvillians [\eq{eq:L_split}] and the projection super-operator [\eq{eq:P_def}], and insert the explicit forms of $\hat{H}_0$ and $\hat{H}_1$ [\eq{eq:H_split}]. The result is a first-order rate equation between the subsystem populations
\begin{equation}\label{eq:FRET_rateeqn}
    \dot{P}_A(t) = \sum_{B \neq A} [ -K_{BA} P_A(t) + K_{AB} P_B(t) ],
\end{equation}
where the rate constant $K_{BA}$ governs the population transfer from $\ket{A}$ to $\ket{B}$, and is expressed as
\begin{equation}\label{eq:FRET_rateconst}
    K_{BA} = \frac{2 |V_{AB}|^2}{\hbar^2} \: \text{Re} \int_0^\infty \tr_\ti{b} [ \hat{U}_B(t') \hat{R}_A \hat{U}_A^\dagger(t') ] \: dt'.
\end{equation}
In the above, $\hat{U}_A(t')$ is the shorthand notation for the unitary operator
\begin{equation}\label{eq:u_t}
    \hat{U}_A (t') = \exp \bigg( -\frac{it'\hat{h}_A}{\hbar} \bigg).
\end{equation}
For brevity, from now on the dependence on $t'$ of the operators $\{ \hat{U}_A(t') \}$ will be omitted.

Equations~(\ref{eq:FRET_rateeqn})--(\ref{eq:u_t}) are valid for general types of bath and subsystem-bath interaction, as we did not make any assumption about the specific form of $\hat{H}_\ti{bath}$ and $\hat{H}_\ti{int}$ up to this point. The calculation of the rate constants $\{K_{AB}\}$ is possible if the trace of the operator product $\tr_\ti{b} [ \hat{U}_B \hat{R}_A \hat{U}_A^\dagger ]$ converges to zero rapidly enough as $t'$ increases, so that the value of the integral in \eq{eq:FRET_rateconst} is well-defined. For relatively simple bath models, such as linearly coupled harmonic oscillator or weakly coupled $\frac{1}{2}$-spins, it is possible to obtain analytical expressions of the trace as will be illustrated in \stn{section:specific}. Even in the situations in which the analytical expression cannot be obtained, it will be still possible to evaluate the trace by factorizing it into the quantities arising from individual bath components and evaluating each of them numerically. Namely, the equilibrium bath densities $\{\hat{R}_A\}$ can be expressed as $\hat{R}_A = \prod_j \hat{r}_{Aj}$, where $\hat{r}_{Aj}$ is the thermal density operator for a single bath mode
\begin{equation}\label{eq:r_onemode}
    \hat{r}_{Aj} = \frac{\exp(-\beta \hat{v}_{Aj})}{\tr_j [ \exp(-\beta \hat{v}_{Aj}) ]},
\end{equation}
with $\tr_j$ being the trace over the subspace spanned by the $j$-th bath component. Then we numerically calculate the traces of the operator products arising from the individual bath components
\begin{equation}\label{eq:trace_indiv}
    \tr_j [ \hat{u}_{Bj} \hat{r}_{Aj} \hat{u}_{Aj}^\dagger] = \tr_j \bigg[ \exp \bigg( \frac{it' \hat{v}_{Bj}}{\hbar} \bigg) \hat{r}_{Aj} \exp \bigg( - \frac{it' \hat{v}_{Aj}}{\hbar} \bigg) \bigg],
\end{equation}
where the unitary operators $\{ \hat{u}_{Aj} \}$ are defined in an analogous manner to \eq{eq:u_t}. Practically, \eq{eq:trace_indiv} is calculated by representing the operators in a trace by using a finite number of bath quantum states that faithfully represent $\hat{r}_{Aj}$ with a desired accuracy. After repeating such a procedure for all bath components, the full trace can be constructed according to
\begin{equation}\label{eq:pop_trace_prod}
    \tr_\ti{b} [ \hat{U}_B \hat{R}_A \hat{U}_A^\dagger ] = \exp \bigg( -\frac{it'(E_B - E_A)}{\hbar} \bigg) \prod_j \tr_j [ \hat{u}_{Bj} \hat{r}_{Aj} \hat{u}_{Aj}^\dagger],
\end{equation}
which is derived from \eq{eq:h_proj}.

Before we move onto the calculation of dissipation, we note that \eq{eq:FRET_rateeqn} is an ``incoherent'' master equation which is only dependent on the state population and does not account for any coherences between the subsystem states. This is because we are only focusing on the dynamics of the projected density $\hmc{P}\hat{\rho}(t)$, and discarded the feedback from the bath onto the subsystem by making the second-order approximation in \eq{eq:2nd_order}. We note, however, that a recent work\cite{Trushechkin2019} illustrated a way to simultaneously follow the dynamics of both $\hmc{P}\hat{\rho}(t)$ and $\hmc{Q}\hat{\rho}(t)$, while also rigorously addressing how the dynamics of the subsystem is affected by the state of the bath.

\subsection{Quantifying the dissipation by individual bath modes using projection operator technique}\label{subsection:calc_diss}
The dissipation accounts for the transfer of energy from the subsystem to the surroundings, which is reflected in how the energy expectation value regarding the bath-related component $\langle \hat{H}_\ti{bath} + \hat{H}_\ti{int} \rangle$ changes with time. 
If we decompose this expectation value according to \eq{eq:bath_component}, the energy of the $j$-th bath component is expressed by $E_j (t) = \tr[\hat{h}_j \hat{\rho}(t)]$, whose time-derivative 
\begin{equation}\label{eq:diss_eq}
    \dot{E}_j (t) = \tr \bigg[ \hat{h}_j \frac{d}{dt} \hat{\rho}(t) \bigg]
\end{equation}
becomes the rate of dissipation into this specific bath component. Our goal is to derive a practical expression for \eq{eq:diss_eq} using the projection operator technique. In this paper, we aim to calculate the amount of dissipation solely arising from the dynamics of $\hmc{P}\hat{\rho}(t)$. Then, the simplest approach would be to substitute $\hat{\rho}(t)$ on the right-hand side by $\hmc{P}\hat{\rho}(t)$,
\begin{equation}\label{eq:diss_naive}
    \dot{E}_j (t) \overset{?}{=} \tr \bigg[ \hat{h}_j \frac{d}{dt} [ \hmc{P}\hat{\rho}(t) ] \bigg].
\end{equation}
However, it turns out that \eq{eq:diss_naive} is not sufficient to capture the dissipation because $\hmc{P}$ [\eq{eq:P_def}] does not allow any change in the bath density. This constraint becomes the most detrimental when the equilibrium energy of the bath component becomes identical for all subsystem states, as in the linearly coupled harmonic oscillator bath model (\stn{subsection:ho}) which is often used to study open quantum systems. We can explicitly show this by replacing $\hmc{P}\hat{\rho}(t)$ with the right-hand side of \eq{eq:P_def},
\begin{equation}\label{eq:naive}
    \tr\bigg[ \hat{h}_j \frac{d}{dt} [\hmc{P} \hat{\rho}(t)] \bigg] = \sum_A \bigg( \frac{d P_A (t)}{dt} \tr_\ti{b} [\hat{v}_{Aj} \hat{R}_A] \bigg),
\end{equation}
and then factoring the bath energy $\epsilon_j = \tr_\ti{b} [\hat{v}_{Aj} \hat{R}_A]$, which we assumed to be state-independent, out from the right-hand side of \eq{eq:naive} to yield
\begin{equation}\label{eq:no_diss}
    \tr\bigg[ \hat{h}_j \frac{d}{dt} [\hmc{P} \hat{\rho}(t)] \bigg] = \epsilon_j \bigg[ \frac{d}{dt} \bigg( \sum_A P_A(t) \bigg) \bigg] = 0.
\end{equation}
In the last equality, we used the property that the sum of the electronic populations $\sum_A P_A(t)$ is always conserved throughout the dynamics. The fact that \eq{eq:diss_naive} yields a vanishing dissipation for a widely used model of open quantum system dynamics states that we need to devise a better method to calculate the dissipated energy.

To get a physically meaningful dissipation by a bath component, we must allow its energy to change while also being consistent with the population dynamics governed by \eq{eq:FRET_rateeqn}. We suggest that such requirements can be fulfilled by factorizing the overall projection operator into $\hmc{P} = \hat{p}_j \hmc{P}_{j-}$ where $\hat{p}_j$ acts on the $j$-th bath component and $\hmc{P}_{j-}$ on all others:
\begin{subequations}\label{eq:proj_factorize}
\begin{equation}
    \hat{p}_j \hat{\rho} = \sum_A \bigg( \tr_j [\bra{A} \hat{\rho} \ket{A} ] \ket{A} \bra{A} \otimes \hat{r}_{Aj} \bigg),
\end{equation}
\begin{equation}
    \hmc{P}_{j-} \hat{\rho} = \sum_A \bigg( \tr_{\ti{b}, j-} [\bra{A} \hat{\rho} \ket{A} ] \ket{A} \bra{A} \otimes \hat{R}_{A, j-} \bigg).
\end{equation}
\end{subequations}
In Eq.~(\ref{eq:proj_factorize}b), we have introduced the equilibrium density operator
\begin{equation}\label{eq:r_rest}
    \hat{R}_{A, j-} = \prod_{k \neq j} \hat{r}_{Ak}
\end{equation}
and the trace $\tr_{\ti{b}, j-}$ over the subspace spanned by all bath modes except the $j$th component. Naturally, $\hat{R}_A = \hat{R}_{A, j-} \hat{r}_{Aj}$ and $\tr_{\ti{b}} [ \hat{O} ] = \tr_{\ti{b}, j-} \big[ \tr_j [\hat{O}] \big]$ are satisfied for an arbitrary operator $\hat{O}$.

In \app{section:equivalence}, we show that
\begin{equation}\label{eq:equivalence}
    \dot{E}_j (t) = \tr \bigg[ \hat{h}_j \frac{d}{dt} [ \hmc{P}_{j-} \hat{\rho}(t) ] \bigg]
\end{equation}
is valid under the assumption $\hat{\rho}(t) = \hmc{P}\hat{\rho}(t)$ [\eq{eq:P_def}], which reflects that we are trying to calculate the dissipation specifically induced by the evolution of $\hmc{P}\hat{\rho}(t)$. \Eq{eq:equivalence} lets us capture the amount of energy dissipated into a bath component before it is quenched by the remaining part of the projection operator $\hat{p}_{j}$. What now remains is converting \eq{eq:equivalence} to a practical expression, which can be achieved by following a procedure analogous to \stn{subsection:Hamiltonian}. Namely, as $\hmc{P}_{j-}^2 = \hmc{P}_{j-}$, we can replace $\hmc{P}$ in \eq{eq:exact_solution} by $\hmc{P}_{j-}$ and obtain
\begin{equation}\label{eq:exact_solution_split}
    \begin{split}
    \frac{d}{dt} [\hmc{P}_{j-} \hat{\rho}(t)] &= -\frac{i}{\hbar} \hmc{P}_{j-} \hmc{L} \hmc{P}_{j-} \hat{\rho}(t) \\
    &-\frac{1}{\hbar^2} \int_0^t \hmc{P}_{j-} \hmc{L} \hmc{Q}_{j-} \exp \bigg[ -\frac{i(t - t')}{\hbar} \hmc{Q}_{j-} \hmc{L} \bigg] \hmc{Q}_{j-} \hmc{L} \hmc{P}_{j-} \hat{\rho}(t') \: dt',
    \end{split}
\end{equation}
where $\hmc{Q}_{j-} = \hmc{I} - \hmc{P}_{j-}$. The next step is expressing $\hmc{Q}_{j-}$ in \eq{eq:exact_solution_split} in terms of $\hmc{P}_{j-}$ and making further simplifications by using $\hmc{P}_{j-} \hmc{L}_0 \hat{\rho} = \hmc{L}_0 \hmc{P}_{j-} \hat{\rho} = \hmc{P}_{j-} \hmc{L}_1 \hmc{P}_{j-} \hat{\rho} = 0$, which is satisfied when $\hat{\rho} = \hmc{P}\hat{\rho}$. Finally, by employing the Markov approximation, we arrive at an expression identical to \eq{eq:rate_prim} except $\hmc{P}$ is replaced by $\hmc{P}_{j-}$,
\begin{equation}\label{eq:dproj_m_rho_dt}
    \frac{d}{dt} [\hmc{P}_{j-} \hat{\rho}(t)] = -\frac{1}{\hbar^2} \int_0^\infty \hmc{P}_{j-} \hmc{L}_1 \exp(-i t' \hmc{L}_0 / \hbar) \hmc{L}_1 \hmc{P}_{j-} \hat{\rho}(t) \: dt'.
\end{equation}
Expanding the integrand leads us to
\begin{equation}\label{eq:integrand}
    \begin{split}
    & \hmc{P}_{j-} \hmc{L}_1 \exp(-i t' \hmc{L}_0 / \hbar) \hmc{L}_1 \hmc{P}_{j-} \hat{\rho}(t) = \sum_A \sum_{B \neq A} \bigg[ |V_{AB}|^2 \ket{A} \bra{A} \\
    &\otimes \bigg( P_A(t) \tr_{\ti{b}, j-} [ \hat{U}_B \hat{R}_A \hat{U}_A^\dagger ] - P_B(t) \tr_{\ti{b}, j-} [ \hat{U}_B \hat{R}_B \hat{U}_A^\dagger 
    ] \bigg) \otimes \hat{R}_{A, j-} \bigg] + \ti{H.c.}
    \end{split}
\end{equation}
Note that the partial trace objects in \eq{eq:integrand} are operators in the subspace spanned by the $j$th component, in contrast to the scalar quantities in \eq{eq:FRET_rateconst}. To proceed, we factorize the bath operators in the traces into contributions from the $j$th component and the rest. For $\{ \hat{R}_A \}$ we can recognize \eqs{eq:r_onemode}{eq:r_rest}, and for $\{ \hat{U}_A \}$ [\eq{eq:u_t}] we have \eq{eq:trace_indiv} and
\begin{equation}\label{eq:u_rest}
    \hat{U}_{A,j-} = \exp \bigg[ -\frac{it'}{\hbar} \bigg( E_A + \sum_{k \neq j} \hat{v}_{Ak} \bigg) \bigg],
\end{equation}
so that $\hat{U}_A = \hat{u}_{Aj} \hat{U}_{A,j-}$. We now plug these factorized bath operators in \eq{eq:integrand} and use the resulting expression with \eq{eq:dproj_m_rho_dt} to expand \eq{eq:equivalence}. At the end, we obtain the rate equation for the dissipation
\begin{equation}\label{eq:diss_eqn}
    \dot{E}_{j} (t) = \sum_A \sum_{B < A} [ \mc{K}_{BA}^j P_A(t) + \mc{K}_{AB}^j P_B (t) ],
\end{equation}
whose rate constants $\{ \mc{K}_{BA}^j \}$ are given by
\begin{equation}\label{eq:diss_const}
    \mc{K}_{BA}^j = \frac{2 |V_{AB}|^2}{\hbar^2} \: \ti{Re} \int_0^\infty \tr_{\ti{b}, j-} [ \hat{U}_{B, j-} \hat{R}_{A, j-} \hat{U}_{A, j-}^\dagger ] \tr_j [ (\hat{v}_{Bj} - \hat{v}_{Aj}) \hat{u}_{Bj} \hat{r}_{Aj} \hat{u}_{Aj}^\dagger ] \: dt'.
\end{equation}
The integrand of \eq{eq:diss_const} can be evaluated by following a similar procedure as that of \eq{eq:FRET_rateconst}, for general types of the bath and subsystem-bath interaction. If an analytical expression for the trace $\tr_j[\hat{u}_{Bj} \hat{r}_{Aj} \hat{u}_{Aj}^\dagger]$ is available, we can utilize the relation
\begin{equation}\label{eq:tderiv_trace}
    \tr_j [ (\hat{v}_{Bj} - \hat{v}_{Aj}) \hat{u}_{Bj} \hat{r}_{Aj} \hat{u}_{Aj}^\dagger ] = i \hbar \frac{d}{dt'} \tr_j [\hat{u}_{Bj} \hat{r}_{Aj} \hat{u}_{Aj}^\dagger ].
\end{equation}
Eqs.~(\ref{eq:diss_eqn})--(\ref{eq:tderiv_trace}) summarize the main findings of this work, which provide an efficient framework for calculating dissipation underlying Markovian quantum master equation (MQME). For the remaining part of the paper, the developed method will be referred to as MQME-D to highlight our extension of MQME toward the analysis of dissipation.

We emphasize that the usage of the factorized projection operator $\hmc{P}_{j-}$ [\eq{eq:equivalence}] by MQME-D does not affect the population dynamics governed by the original MQME [\eq{eq:FRET_rateeqn}]. It can be thought that at every time instance $t$, we start from the projected density $\hmc{P} \hat{\rho} (t)$ and temporarily lift the projection for the $j$-th bath component. The density matrix is then propagated for an infinitesimal amount of time $dt$, and the dissipated energy into the $j$-th bath component is evaluated based on \eq{eq:equivalence}. Immediately after that, the remaining part of the projection operator $\hat{p}_j$ is applied to the intermediate density $\hmc{P}_{j-}\hat{\rho}(t+dt)$ and quenches the excess energy of the $j$-th component gained by the dissipation. By doing so, the density returns again to the fully projected form $\hmc{P}\hat{\rho}(t+dt)$, in agreement with the evolution according to \eq{eq:FRET_rateeqn}. The procedure just described can be rationalized by the identity
\begin{equation}\label{eq:equiv_dynamics}
    \frac{d}{dt} [ \hmc{P}\hat{\rho}(t) ] = \hat{p}_j \frac{d}{dt} \big[ \hmc{P}_{j-} \hat{\rho}(t) \big],
\end{equation}
whose proof is outlined in \app{section:proof_equiv}.

\subsection{Proof of thermodynamic principles}\label{subsection:proof}
We now confirm that the dissipation calculated by MQME-D [\eq{eq:diss_eqn}] satisfies thermodynamic principles, namely energy conservation and detailed balance.

\subsubsection{Energy conservation}
We first prove the energy conservation, which states that the rate of energy loss from the subsystem must be equal to the rate of energy gain by the entire bath. This is expressed as
\begin{equation}\label{eq:econserv}
    \dot{E}_\ti{sub}(t) + \sum_j \dot{E}_j (t) = 0,
\end{equation}
where $E_\ti{sub}(t)$ is the expectation value for the energy of the subsystem
\begin{equation}\label{eq:E_sys}
    E_\ti{sub}(t) = \tr[\hat{H}_\ti{sub} \: \hat{\rho}(t)].
\end{equation}
Even though the calculation of the dissipation is based on \eq{eq:equivalence} which is exact, it would be still meaningful to check whether \eq{eq:econserv} is still valid despite the Markov approximation applied in the derivation. It is also yet another demonstration of the consistency between the dynamics of the population [\eq{eq:FRET_rateeqn}] and dissipation [\eq{eq:diss_eqn}], apart from \app{section:proof_equiv}. To begin with, we combine \eq{eq:E_sys} with \eq{eq:P_def} by recalling that we are assuming that the identity $\hat{\rho}(t) = \hmc{P} \hat{\rho}(t)$ is satisfied at every instance (\stn{subsection:calc_diss}), and therefore
\begin{equation}\label{eq:Eel}
    E_\ti{sub}(t) = \sum_A E_A P_A(t).
\end{equation}
Taking the time-derivative of both sides and expanding $\dot{P}_A(t)$ with \eqs{eq:FRET_rateeqn}{eq:FRET_rateconst} lead us to
\begin{equation}\label{eq:Eel_deriv}
    \begin{split}
    \dot{E}_\ti{sub}(t)
    &= -\frac{2}{\hbar^2} \sum_A \sum_{B \neq A} \bigg[ E_A |V_{AB}|^2 \: \ti{Re} \bigg( P_A(t) \int_0^\infty \tr_\ti{b} [ \hat{U}_{B} \hat{R}_A \hat{U}_{A}^\dagger ] \: dt' \\
    &- P_B(t)  \int_0^\infty \tr_\ti{b} [ \hat{U}_{A} \hat{R}_B \hat{U}_{B}^\dagger ] \: dt' \bigg) \bigg].
    \end{split}
\end{equation}
When this is combined with \eqs{eq:diss_eqn}{eq:diss_const}, we get
\begin{equation}\label{eq:econserv_exp}
    \begin{split}
    &\dot{E}_\ti{sub}(t) + \sum_j \dot{E}_j (t) \\
    &= - \frac{2}{\hbar^2} \sum_A \sum_{B < A} \bigg[ |V_{AB}|^2 \: \ti{Re} \bigg( P_A(t) \int_0^\infty \tr_\ti{b} [(\hat{h}_A - \hat{h}_B) \hat{U}_B \hat{R}_A \hat{U}_A^\dagger] \: dt' \\
    &+ P_B(t) \int_0^\infty \tr_\ti{b} [(\hat{h}_B - \hat{h}_A) \hat{U}_A \hat{R}_B \hat{U}_B^\dagger] \: dt' \bigg) \bigg],
    \end{split}
\end{equation}
where we have used the definition of $\hat{h}_A$ [\eq{eq:h_proj}] to condense the bath operators. Then, by recognizing the identity
\begin{equation}
    \tr_\ti{b} [(\hat{h}_A - \hat{h}_B) \hat{U}_B \hat{R}_A \hat{U}_A^\dagger] = - i\hbar \frac{d}{dt'} \tr_\ti{b} [\hat{U}_B \hat{R}_A \hat{U}_A^\dagger],
\end{equation}
we can perform the integration over $t'$ to derive
\begin{equation}\label{eq:int_vib}
    \int_0^\infty \tr_\ti{b} [(\hat{h}_A - \hat{h}_B) \hat{U}_B \hat{R}_A \hat{U}_A^\dagger] \: dt' = i\hbar \bigg( 1 - \lim_{t' \rightarrow \infty} \tr_\ti{b} [\hat{U}_B \hat{R}_A \hat{U}_A^\dagger] \bigg).
\end{equation}
\Eq{eq:int_vib} shows that the integrals in \eq{eq:econserv_exp} yield purely imaginary numbers if $\tr_\ti{b} [\hat{U}_B \hat{R}_A \hat{U}_A^\dagger]$ decays to zero as $t' \rightarrow \infty$, eventually converting \eq{eq:econserv_exp} to \eq{eq:econserv}. Therefore, as long as the integrals for the rate constants [\eq{eq:FRET_rateconst}] are well-defined, \eq{eq:econserv} is also valid and we can conclude that MQME-D satisfies the energy conservation.

\subsubsection{Detailed balance}
At the steady state, the ratio between the rate constants for population transfer in the opposite directions must satisfy
\begin{equation}\label{eq:pop_detbal}
    \frac{K_{AB}}{K_{BA}} = \frac{P_A(\infty)}{P_B(\infty)},
\end{equation}
which arises from the connection of \eq{eq:FRET_rateeqn} to the steady-state condition, $\dot{P}_A(\infty) = 0$ for all $A$. The net dissipation into every bath component must also vanish at this point, so that $\lim_{t \rightarrow \infty} \dot{E}_j(t) = 0$. Combining this condition with \eqs{eq:diss_eqn}{eq:pop_detbal} gives us a relation between dissipation rate constants in the opposite directions,
\begin{equation}\label{eq:diss_detbal}
    - \frac{\mc{K}_{AB}^j}{\mc{K}_{BA}^j} = \frac{K_{AB}}{K_{BA}},
\end{equation}
which must be satisfied in order to maintain its consistency with the population dynamics. To prove \eq{eq:diss_detbal}, we first need to derive the detailed balance condition for the state populations, which is achieved by following an approach motivated by Ref.~\citenum{Banchi2013}. We start by defining a function in the time-domain
\begin{equation}\label{eq:tprof0}
    f_{BA}(t') = \exp \bigg( \frac{it'}{\hbar} (E_B - E_A ) \bigg) \tr_\ti{b}[ \hat{U}_B \hat{R}_A \hat{U}_A^\dagger ],
\end{equation}
and perform the Wick rotation $t' \rightarrow t' - i\hbar \beta$ to get
\begin{equation}\label{eq:tprof_interm0}
    \begin{split}
    f_{BA} (t' - i \hbar \beta) &= \exp \bigg[ \bigg( \frac{it'}{\hbar} + \beta \bigg) (E_B - E_A) \bigg] \\
    &\times \tr_\ti{b} [ \hat{U}_B \exp(-\beta \hat{h}_B) \hat{R}_A \exp(\beta \hat{h}_A) \hat{U}_A^\dagger ].
    \end{split}
\end{equation}
We now introduce the identity
\begin{equation}\label{eq:eqdens_relation}
    \exp(-\beta\hat{h}_B) \hat{R}_A \exp(\beta \hat{h}_A) = \frac{\tr_\ti{b} [\exp(- \beta \hat{h}_B)]}{\tr_\ti{b} [\exp(- \beta \hat{h}_A)]} \hat{R}_B,
\end{equation}
which is immediately validated by replacing $\hat{R}_A$ and $\hat{R}_B$ with its definition [\eq{eq:R_def}]. If we insert \eq{eq:eqdens_relation} in \eq{eq:tprof_interm0} and take the complex conjugate, we can elucidate
\begin{equation}\label{eq:epsilon_eq01}
    f_{BA} (t' - i \hbar \beta) = \frac{\tr_\ti{b} [\exp(-\beta \sum_j \hat{v}_{Bj})]}{\tr_\ti{b} [\exp(-\beta \sum_j \hat{v}_{Aj})]} [f_{AB}(t')]^*,
\end{equation}
where the state energies $E_A$ and $E_B$ were eliminated from the expression by recalling \eq{eq:h_proj}. \Eq{eq:epsilon_eq01} conjoins the two time profiles related to the population transfers in opposite directions. Meanwhile, $f_{BA} (t')$ and its Fourier transform $\tilde{f}_{BA} (\omega)$ are connected by
\begin{equation}\label{eq:epsilon_eq02}
    f_{BA} (t') = \frac{1}{2 \pi} \int_{-\infty}^\infty \tilde{f}_{BA} (\omega) e^{i \omega t'} \: d\omega,
\end{equation}
which can be combined with \eq{eq:epsilon_eq01} to give
\begin{equation}\label{eq:epsilon_eq03}
    f_{AB} (t') = \frac{1}{2\pi} \frac{\tr_\ti{b} [\exp(-\beta \sum_j \hat{v}_{Aj})]}{\tr_\ti{b} [\exp(-\beta \sum_j \hat{v}_{Bj})]} \int_{-\infty}^\infty \tilde{f}_{BA} (\omega) e^{\beta \hbar \omega} e^{-i \omega t'} \: d \omega,
\end{equation}
where we have recognized that $f_{BA}(t') = [f_{BA}(-t')]^*$ which makes $\tilde{f}_{BA}(\omega)$ a real-valued function. We now express the integrand in the explicit expression of $K_{BA}$ [\eq{eq:FRET_rateconst}] in terms of $f_{BA}(t')$ [\eq{eq:tprof0}] and change the lower limit of the integral to $-\infty$ by using the identity $2 \: \ti{Re} \int_0^{\infty} y(z) \: dz = \int_{-\infty}^\infty y(z) \: dz$ which holds when $[y(z)]^* = y(-z)$. Finally, invoking \eqs{eq:epsilon_eq02}{eq:epsilon_eq03} and then carrying out the integrations lead us to
\begin{subequations}\label{eq:epsilon_eq04}
    \begin{equation}
    \begin{split}
    K_{BA} = \frac{|V_{AB}|^2}{\hbar^2} \: \tilde{f}_{BA} \bigg( \frac{E_B - E_A}{\hbar} \bigg),
    \end{split}
    \end{equation}
    \begin{equation}
    \begin{split}
    K_{AB} = \frac{|V_{AB}|^2}{\hbar^2} \frac{\tr_\ti{b} [\exp(-\beta \hat{h}_A)]}{\tr_\ti{b} [\exp(-\beta \hat{h}_B)]} \tilde{f}_{BA} \bigg( \frac{E_B - E_A}{\hbar} \bigg),
    \end{split}
    \end{equation}
\end{subequations}
which gives the simplified ratio between the rate constants as
\begin{equation}\label{eq:detbal_op}
    \frac{K_{AB}}{K_{BA}} = \frac{\tr_\ti{b} [\exp(-\beta \hat{h}_A)]}{\tr_\ti{b} [\exp(-\beta \hat{h}_B)]}.
\end{equation}
\Eq{eq:detbal_op} is the detailed balance condition which shows that the ratio between the steady-state populations [\eq{eq:pop_detbal}] follows the Boltzmann distribution dictated by the PESs associated with the two relevant subsystem states. When the two PESs have an identical landscape as in the Hamiltonian models discussed in \stn{section:specific}, we have $\tr_\ti{b} [\exp(-\beta \sum_j \hat{v}_{Aj})] = \tr_\ti{b} [\exp(-\beta \sum_j \hat{v}_{Bj})]$ and the ratio between the traces in \eq{eq:detbal_op} further simplifies into $\exp[\beta(E_B - E_A)]$.

We now prove that the detailed balance is also satisfied for the dissipation. For this purpose, we re-write the intermediate expression for the dissipation rate constant [\eq{eq:diss_const}] by combining the two traces,
\begin{equation}\label{eq:Kdiss_operator}
    \mc{K}_{BA}^j = \frac{2|V_{AB}|^2}{\hbar^2} \: \ti{Re} \int_0^\infty \tr_\ti{b}[ (\hat{v}_{Bj} - \hat{v}_{Aj}) \hat{U}_B \hat{R}_A \hat{U}_A^\dagger ] \: dt',
\end{equation}
which is allowed by the fact that $\tr[\hat{O}]\tr[\hat{O}'] = \tr[\hat{O} \otimes \hat{O}']$ for arbitrary operators $\hat{O}$ and $\hat{O}'$ acting on orthogonal subspaces. \Eq{eq:Kdiss_operator} now resembles \eq{eq:FRET_rateconst} except for the additional operator $\hat{v}_{Bj} - \hat{v}_{Aj}$ in the trace, which prompts us to follow the similar procedure as we carried out for the state populations. We thus define a time profile analogous to \eq{eq:tprof0},
\begin{equation}\label{eq:tprof}
    \mc{T}_{BA}^j(t') = \exp \bigg( \frac{it'}{\hbar} (E_B - E_A ) \bigg) \tr_\ti{b}[ (\hat{v}_{Bj} - \hat{v}_{Aj}) \hat{U}_B \hat{R}_A \hat{U}_A^\dagger ],
\end{equation}
and perform the Wick rotation to arrive at
\begin{equation}\label{eq:tprof_interm}
    \begin{split}
    \mc{T}_{BA}^j (t' - i \hbar \beta) &= \exp \bigg[ \bigg( \frac{it'}{\hbar} + \beta \bigg) (E_B - E_A) \bigg] \\
    &\times \tr_\ti{b} [ (\hat{v}_{Bj} - \hat{v}_{Aj}) \hat{U}_B \exp(-\beta \hat{h}_B) \hat{R}_A \exp(\beta \hat{h}_A) \hat{U}_A^\dagger ].
    \end{split}
\end{equation}
By invoking \eq{eq:eqdens_relation} and relating the time profiles in the opposite directions, we obtain the relation
\begin{equation}\label{eq:epsilon_eq1}
    \mc{T}_{BA}^j (t' - i \hbar \beta) = - [\mc{T}_{AB}^j(t')]^*,
\end{equation}
which, in contrast to \eq{eq:epsilon_eq01}, has an additional negative sign on the right-hand side arising from $\hat{v}_{Bj} - \hat{v}_{Aj}$ in the traces of \eqs{eq:tprof}{eq:tprof_interm}. Continuing along the steps we took for the state populations [Eqs.~(\ref{eq:epsilon_eq02})--(\ref{eq:detbal_op})] eventually leads us to
\begin{equation}\label{eq:detbal_op_diss}
    \frac{\mc{K}_{AB}^j}{\mc{K}_{BA}^j} = -\frac{\tr_\ti{b} [\exp(-\beta \hat{h}_A)]}{\tr_\ti{b} [\exp(-\beta \hat{h}_B)]},
\end{equation}
from which we can straightforwardly prove \eq{eq:diss_detbal} by comparing it with \eq{eq:detbal_op}.

\section{Application to Specific Models}\label{section:specific}
In this section, we apply MQME-D developed in \stn{subsection:calc_diss} to prototypical models of open quantum systems. \Stn{subsection:ho} first deals with a harmonic bath with a general form of linear subsystem-bath coupling, and then shows that the results reduce to the formulae reported in our earlier work.\cite{Kim2021} \Stn{subsection:spin} treats a model which consists of $\frac{1}{2}$-spins and demonstrates that the results recovers the known connection to the harmonic bath model in the weak-coupling limit.

\subsection{Harmonic Oscillator Bath}\label{subsection:ho}
\subsubsection{General Linear Subsystem-Bath Coupling}
We first apply MQME-D to the bath composed of harmonic oscillators that are linearly coupled to the subsystem states, which is represented by the Hamiltonian
\begin{equation}\label{eq:H_bath_ho}
    \hat{H}_\ti{bath} = \sum_j \bigg(\frac{\hat{p}_j^2}{2} +  \frac{\omega_j^2 \hat{x}_j^2}{2} \bigg),
\end{equation}
\begin{equation}\label{eq:H_int_ho}
    \hat{H}_\ti{int} = \sum_A \bigg[ \ket{A} \bra{A} \otimes \sum_j \bigg(- \omega_j^2 d_{Aj} \hat{x}_j + \frac{\omega_j^2 d_{Aj}^2}{2} \bigg) \bigg].
\end{equation}
In the equations above, $\hat{p}_j$ and $\hat{x}_j$ are the mass-weighted momentum and position operators for the $j$-th bath mode, and $d_{Aj}$ determines the strength of interaction between $\ket{A}$ and the $j$-th bath mode. This type of Hamiltonian is used to describe a variety of chemical phenomena involving vibronic interactions, such as excitation energy transfer,\cite{Lei2020,Cho2021} charge transfer,\cite{Song1993,Evans1996} and singlet fission.\cite{Morrison2017,Masayoshi2019} In this model, $\hat{v}_{Aj}$ is given by
\begin{equation}
    \hat{v}_{Aj} = \frac{\hat{p}_j^2}{2} + \frac{\omega_j^2}{2} (\hat{x}_j - d_{Aj})^2,
\end{equation}
whose energy expectation value can be calculated based on \eqs{eq:R_def}{eq:h_proj}, yielding
\begin{equation}\label{eq:ener_expval}
    \epsilon_j = \tr_\ti{b} [\hat{v}_{Aj} \hat{R}_A] = \frac{\hbar \omega_j}{2} \coth \bigg( \frac{\beta \hbar \omega_j}{2} \bigg).
\end{equation}
\Eq{eq:ener_expval} shows that, for the model depicted by \eqs{eq:H_bath_ho}{eq:H_int_ho}, the thermal equilibrium energy of a bath component is indeed independent of the subsystem states, as supposed by \eq{eq:no_diss}. As we have alluded in \stn{subsection:calc_diss}, the na{\"i}vely constructed expression [\eq{eq:diss_naive}] would yield zero dissipation under such a circumstance, which highlights the importance of \eqs{eq:diss_eqn}{eq:diss_const} for quantitative decomposition of the dissipated energy.

The energy difference between the zero-phonon excitation energy and the Franck-Condon excitation energy at the origin of the bath coordinates is known as the reorganization energy. We introduce the notation
\begin{equation}\label{eq:reorg}
    \Lambda_{AB} = \sum_j \frac{\omega_j^2 d_{Aj} d_{Bj}}{2},
\end{equation}
such that the reorganization energy for $\ket{A}$ becomes $\Lambda_{AA}$. We also define the bath spectral density (BSD) which is the profile of subsystem-bath interaction strength in the frequency domain,
\begin{equation}\label{eq:BSD}
    J_{AB}(\omega) = \sum_j \frac{\omega_j^3 d_{Aj} d_{Bj}}{2} \delta(\omega - \omega_j).
\end{equation}

The traces of the operator products required to evaluate the dissipation rate constants can be expressed analytically:
\begin{equation}\label{eq:product_three}
    \begin{split}
    \tr_j[\hat{u}_{Bj} \hat{r}_{Aj} \hat{u}_{Aj}^\dagger] & =\exp \bigg[ -\frac{\omega_j^2 (d_{Bj} - d_{Aj})^2}{2 \hbar} \\
    & \times \bigg\{ \coth \bigg( \frac{\beta \hbar \omega_j}{2} \bigg) \frac{ 1 - \cos(\omega_j t') }{\omega_j} + i \frac{\sin (\omega_j t')}{\omega_j} \bigg\} \bigg],
    \end{split}
\end{equation}
\begin{equation}\label{eq:product_four}
    \begin{split}
    \tr_j [ (\hat{v}_{Bj} - \hat{v}_{Aj}) \hat{u}_{Bj} \hat{r}_{Aj} \hat{u}_{Aj}^\dagger ] & = \frac{\omega_j^2 (d_{Bj} - d_{Aj})^2}{2 \hbar} \tr_j[\hat{u}_{Bj} \hat{r}_{Aj} \hat{u}_{Aj}^\dagger] \\
    &\times \bigg[ \cos(\omega_j t') - i \coth \bigg( \frac{\beta \hbar \omega_j}{2} \bigg) \sin(\omega_j t') \bigg].
    \end{split}
\end{equation}
\Eq{eq:product_three} can be derived by either invoking the polaron transformation\cite{Jang2002,Jang2012} or switching to the interaction picture and applying the cumulant expansion.\cite{Mukamel1983,Sung2001} However, one should be aware that the cumulant expansion must be used with care in this case, due to the emergence of the time-ordered integral.\cite{Fox1976} \Eq{eq:product_four} straightforwardly emanates from \eq{eq:product_three} via \eq{eq:tderiv_trace}. These results can be inserted in \eq{eq:diss_const} to obtain the expression for the dissipation rate constants,
\begin{equation}\label{eq:diss_const2}
    \begin{split}
    \mc{K}_{BA}^j &= \frac{2 |V_{AB}|^2}{\hbar^2} (\lambda_{AA}^j - 2 \lambda_{AB}^j + \lambda_{BB}^j) \\
    &\times \ti{Re} \int_0^\infty \exp \bigg( - \frac{it' (E_B - E_A + \Lambda_{AA} - 2 \Lambda_{AB} + \Lambda_{BB})}{\hbar} \bigg) \\
    &\times \exp [- g_{AA}(t') + 2 g_{AB}(t') - g_{BB}(t') ] \\
    &\times \bigg[ \cos(\omega_j t') - i\coth \bigg( \frac{\beta \hbar \omega_j}{2} \bigg) \sin(\omega_j t') \bigg] \: dt',
    \end{split}
\end{equation}
where $\lambda_{AB}^j = \omega_j^2 d_{Aj} d_{Bj}/2$ is the generalized reorganization energy associated with the $j$-th bath mode, and $g_{AB}(t')$ is the line-broadening function,
\begin{equation}\label{eq:line_broaden}
    g_{AB}(t') = \frac{1}{\hbar} \int_0^\infty J_{AB}(\omega) \bigg[ \coth \bigg( \frac{\beta \hbar \omega}{2} \bigg) \frac{1 - \cos(\omega t')}{\omega^2} + i \: \frac{\sin(\omega t') - \omega t'}{\omega^2} \bigg] \: d\omega.
\end{equation}
In Paper II, we will conduct a thorough assessment of the accuracy of the dissipation calculated by \eqs{eq:diss_eqn}{eq:diss_const2}.

\subsubsection{Local Coupling}
When each of the harmonic oscillators in the bath is exclusively coupled to a single subsystem state, \eqs{eq:H_bath_ho}{eq:H_int_ho} reduce to the so-called local bath model,
\begin{equation}\label{eq:H_bath_loc}
    \hat{H}_\ti{bath} = \sum_A \sum_j \bigg(\frac{\hat{p}_{Aj}^2}{2} +  \frac{\omega_{Aj}^2 \hat{x}_{Aj}^2}{2} \bigg),
\end{equation}
\begin{equation}\label{eq:H_int_loc}
    \hat{H}_\ti{int} = \sum_A \bigg[ \ket{A} \bra{A} \otimes \sum_j \bigg(- \omega_{Aj}^2 d_{Aj} \hat{x}_{Aj} + \frac{\omega_{Aj}^2 d_{Aj}^2}{2} \bigg) \bigg].
\end{equation}
This model is often used to describe excitation energy transfer among electronically coupled chromophore molecules without any significant inter-molecular vibronic coupling.\cite{Sumi1999,Jang2012,Bennett2018} The dissipation rate equation and rate constants for this model can be conveniently derived from \eqs{eq:diss_eqn}{eq:diss_const2} by imposing the condition $d_{Aj} d_{Bj} = 0$ whenever $A \neq B$, which leads to
\begin{equation}\label{eq:diss_eqn_loc}
    \dot{E}_{Aj} (t) = \sum_{B \neq A} [ \mc{K}_{BA}^{Aj} P_A(t) + \mc{K}_{AB}^{Aj} P_B (t) ],
\end{equation}
and
\begin{subequations}\label{eq:diss_rate_local}
\begin{equation}
    \mc{K}_{BA}^{Aj} = \frac{2 |V_{AB}|^2}{\hbar^2} \lambda_{Aj} \mc{I}_{BA}(\omega_{Aj}),
\end{equation}
\begin{equation}
    \mc{K}_{AB}^{Aj} = \frac{2 |V_{AB}|^2}{\hbar^2} \lambda_{Aj} \mc{I}_{AB}(\omega_{Aj}),
\end{equation}
\end{subequations}
with $\lambda_{Aj} = \omega_{Aj}^2 d_{Aj}^2 / 2$. Here, the ``dissipative potential'' $\mc{I}_{AB}(\omega)$ is defined as
\begin{equation}\label{eq:Idiss}
    \begin{split}
    \mc{I}_{BA} (\omega) &= \ti{Re} \int_0^\infty \mc{F}_A^*(t') \mc{A}_B(t') \\
    &\times \bigg[ \cos(\omega t') - i\coth \bigg( \frac{\beta \hbar \omega}{2} \bigg) \sin(\omega t') \bigg] \: dt',
    \end{split}
\end{equation}
which quantifies the capability of a bath mode to induce dissipation under a unit reorganization energy. The time profiles $\mc{F}_A(t')$ and $\mc{A}_A(t')$ are expressed as
\begin{subequations}\label{eq:FA_t}
    \begin{equation}
    \mc{F}_A(t') = \exp \bigg[ -\frac{it'(E_A - \Lambda_{AA})}{\hbar} - g_{AA}^*(t') \bigg],
    \end{equation}
    \begin{equation}
    \mc{A}_A(t') = \exp \bigg[ -\frac{it'(E_A + \Lambda_{AA})}{\hbar} - g_{AA}(t') \bigg],
    \end{equation}
\end{subequations}
which are the Fourier transform of the linear fluorescence and absorption spectra of molecule $A$ in the model of excitation energy transfer.

Reformulating \eq{eq:diss_rate_local} in terms of BSD [\eq{eq:BSD}] converts the rate constants to continuous functions in the frequency domain
\begin{subequations}\label{eq:DSD}
\begin{equation}
    \mc{J}_{BA}^A(\omega) = \frac{2|V_{AB}|^2}{\hbar^2} \frac{J_{AA}(\omega)}{\omega} \mc{I}_{BA}(\omega),
\end{equation}
\begin{equation}
    \mc{J}_{AB}^A(\omega) = \frac{2|V_{AB}|^2}{\hbar^2} \frac{J_{AA}(\omega)}{\omega} \mc{I}_{AB}(\omega).
\end{equation}
\end{subequations}
For the functions $\mc{J}_{BA}^A(\omega)$ and $\mc{J}_{AB}^A(\omega)$, the superscript denotes the molecule which the bath mode is coupled to, and the subscript the direction of population transfer. \Eq{eq:DSD} is called ``dissipative spectral densities,'' as they determine how the dissipated energy is distributed across the frequency domain. If we replace $\mc{K}_{BA}^{Aj}$ and $\mc{K}_{AB}^{Aj}$ in the dissipation rate equation [\eq{eq:diss_eqn_loc}] by $\mc{J}_{BA}^A(\omega) \: d\omega$ and $\mc{J}_{AB}^A(\omega) \: d\omega$, we get an expression for the rate of dissipation through the frequency window $[\omega, \omega + d\omega]$ at a certain time,
\begin{equation}\label{eq:diss_dens}
    \mc{D}_A(\omega, t) \: d\omega = \sum_{B \neq A} [ \mc{J}_{BA}^A(\omega) P_A(t) + \mc{J}_{AB}^A(\omega) P_B(t) ] \: d\omega.
\end{equation}
Equation (\ref{eq:diss_dens}) is indeed identical to Eq.~(51) in our earlier work,\cite{Kim2021} with minor changes in the notation.

\subsection{Spin Bath}\label{subsection:spin}
As a last example, we consider a situation where a two-level subsystem interacts with the nearby spins. We denote the subsystem states by $\ket{+}$ and $\ket{-}$ and express $\hat{H}_\ti{sub}$ as
\begin{equation}\label{eq:spin_sub}
    \hat{H}_\ti{sub} = \frac{E}{2} ( \ket{+}\bra{+} - \ket{-}\bra{-} ) + V ( \ket{+}\bra{-} + \ket{-}\bra{+} ),
\end{equation}
while the bath consists of $\frac{1}{2}$-spins under the presence of a magnetic field along the $z$-direction
\begin{equation}\label{eq:spin_bath}
    \hat{H}_\ti{bath} = \sum_j \frac{\hbar \omega_j}{2} \hat{\sigma}_{zj},
\end{equation}
where $\{\hat{\sigma}_{xj}, \hat{\sigma}_{yj}, \hat{\sigma}_{zj}\}$ are the Pauli operators for the $j$-th spin. In our model, the spins in the bath couple to the subsystem according to
\begin{equation}\label{eq:spin_int}
    \hat{H}_\ti{int} = ( \ket{+}\bra{+} - \ket{-}\bra{-} ) \otimes \bigg( -\sum_{j=1}^n \hbar \gamma_j \hat{\sigma}_{xj} \bigg),
\end{equation}
with the interaction strength quantified by $\{\gamma_j\}$. The Hamiltonian defined by Eqs.~(\ref{eq:spin_sub})--(\ref{eq:spin_int}) has been thoroughly studied\cite{Caldeira1993,Marques1995,Shao1998} and can be thought as a simplified model of a solid-state qubit undergoing relaxation due to the environmental spins in the matrix. By employing
\begin{equation}
    \hat{v}_{\pm j} = \frac{\hbar \omega_j}{2} \hat{\sigma}_{zj} \mp \hbar \gamma_j \hat{\sigma}_{xj}
\end{equation}
with Eqs.~(\ref{eq:r_onemode}), (\ref{eq:trace_indiv}), and (\ref{eq:tderiv_trace}), the operator traces required for calculating the rate constants are evaluated as
\begin{equation}\label{eq:spin_3prod_trace}
    \begin{split}
    \tr_j [\hat{u}_{\pm j} \hat{r}_{\mp j} \hat{u}_{\mp j}^\dagger] = \cos^2(2 \theta_j) + \sin^2(2 \theta_j) \bigg[ \cos(\tilde{\omega}_j t') - i \tanh \bigg( \frac{\beta \hbar \tilde{\omega}_j}{2} \bigg) \sin(\tilde{\omega}_j t') \bigg],
    \end{split}
\end{equation}
\begin{equation}
    \tr_j [(\hat{v}_{\pm j} - \hat{v}_{\mp j}) \hat{u}_{\pm j} \hat{r}_{\mp j} \hat{u}_{\mp j}^\dagger] = \hbar \tilde{\omega}_j \sin^2(2 \theta_j) \bigg[ \tanh \bigg( \frac{\beta \hbar \tilde{\omega}_j}{2} \bigg) \cos(\tilde{\omega}_j t') -i \sin(\tilde{\omega}_j t') \bigg].
\end{equation}
In the above, $\tilde{\omega}_j = \sqrt{\omega_j^2 + 4\gamma_j^2}$ and $\theta_j = \frac{1}{2} \tan^{-1} (2 \gamma_j/\omega_j)$. If the subsystem-bath coupling is weak compared to the Zeeman splitting so that $\gamma_j \ll \omega_j$ and $\omega_j \sim \tilde{\omega}_j$, we can approximate \eq{eq:spin_3prod_trace} by expanding $\cos(2 \theta_j)$ and $\sin(2 \theta_j)$ and keeping the two lowest-order terms in $\theta_j$ in the resulting expression,
\begin{equation}
    \begin{split}
    \tr_j [\hat{u}_{\mp j} \hat{r}_{\pm j} \hat{u}_{\pm j}^\dagger] &\approx 1 - \frac{4 \gamma_j^2}{\omega_j^2} \bigg[ 1 - \cos(\omega_j t') + i \tanh \bigg( \frac{\beta \hbar \omega_j}{2} \bigg) \sin(\omega_j t') \bigg] \\
    &\approx \exp \bigg[ - \frac{4 \gamma_j^2}{\omega_j^2} \bigg\{ 1 - \cos(\omega_j t') + i \tanh \bigg( \frac{\beta \hbar \omega_j}{2} \bigg) \sin(\omega_j t') \bigg\} \bigg],
    \end{split}
\end{equation}
to which we can apply \eq{eq:tderiv_trace} and get
\begin{equation}
    \begin{split}
    \tr_j [(\hat{v}_{\pm j} - \hat{v}_{\mp j}) \hat{u}_{\mp j} \hat{r}_{\pm j} \hat{u}_{\pm j}^\dagger] & = \frac{4 \hbar \gamma_j^2}{\omega_j} \bigg[ \tanh \bigg( \frac{\beta \hbar \omega_j}{2} \bigg) \cos (\omega_j t') - i \sin(\omega_j t') \bigg] \\
    &\times \tr_j [\hat{u}_{\mp j} \hat{r}_{\pm j} \hat{u}_{\pm j}^\dagger].
    \end{split}
\end{equation}
If we combine these results with \eqs{eq:pop_trace_prod}{eq:diss_const}, we can express the rate constants for the population transfer and dissipation as
\begin{equation}
    K_{\mp \pm} = \frac{2 V^2}{\hbar^2} \: \ti{Re} \int_0^\infty \exp \bigg( \frac{i t' (\pm E - 4 \Lambda_\ti{s})}{\hbar} -4 g_\ti{s}(t') \bigg) \: dt',
\end{equation}
\begin{equation}\label{eq:spin_diss}
    \begin{split}
    \mc{K}_{\mp \pm}^j &= \frac{8 V^2}{\hbar^2} \lambda_\ti{s}^j \: \ti{Re} \int_0^\infty \exp \bigg( \frac{i t' (\pm E - 4 \Lambda_\ti{s})}{\hbar} -4 g_\ti{s}(t') \bigg) \\
    &\times \bigg[ \cos(\omega_j t') - i \coth \bigg( \frac{\beta \hbar \omega_j}{2} \bigg) \sin (\omega_j t') \bigg] \: dt',
    \end{split}
\end{equation}
by defining the spectral density, reorganization energy, and line-broadening function of the spin bath as
\begin{equation}
    J_\ti{s}(\omega) = \sum_j \hbar \gamma_j^2 \delta(\omega - \omega_j),
\end{equation}
\begin{equation}
    \lambda_\ti{s}^j = \frac{\hbar \gamma_j^2}{\omega_j} \tanh \bigg( \frac{\beta \hbar \omega_j}{2} \bigg), \quad \Lambda_\ti{s} = \sum_j \lambda_\ti{s}^j,
\end{equation}
\begin{equation}
    g_\ti{s}(t') = \frac{1}{\hbar} \int_0^\infty J_\ti{s}(\omega) \bigg[ \frac{1 - \cos(\omega t')}{\omega^2} + i \tanh \bigg( \frac{\beta \hbar \omega_j}{2} \bigg) \frac{\sin(\omega t') - \omega t'}{\omega^2} \bigg] \: d\omega,
\end{equation}
respectively. By comparing \eq{eq:spin_diss} with its harmonic oscillator counterpart [\eq{eq:diss_const2}], we can see that the dynamics of the spin bath can be described by a surrogate spin-boson model defined by Eqs.~(\ref{eq:H_bath_ho}), (\ref{eq:spin_sub}), and
\begin{equation}
    \hat{H}_\ti{int} = ( \ket{+}\bra{+} - \ket{-}\bra{-} ) \otimes \sum_j \bigg(- \omega_j^2 d_j \hat{x}_j + \frac{\omega_j^2 d_j^2}{2} \bigg),
\end{equation}
whose spectral density is related to that of the spin bath via
\begin{equation}\label{eq:spin_ho_connection}
    J_{\pm \pm}(\omega) = J_\ti{s}(\omega) \tanh \bigg( \frac{\beta \hbar \omega}{2} \bigg).
\end{equation}
We have thus demonstrated that the well-known connection\cite{Schlosshauer2007} between the spin and harmonic oscillator bath in the weak-coupling limit [\eq{eq:spin_ho_connection}] also prevails in the dissipation.

\section{Conclusion}\label{section:Conclusion}

In this paper, we proposed MQME-D, which is a new theoretical framework for decomposing the dissipated energy in open quantum system dynamics into contributions by individual bath components. The developed framework was applied to multiple prototypical Hamiltonian models of open quantum system, which led us to practical expressions that enable efficient decomposition of the dissipation underlying these models. In particular, we showed that MQME-D yields identical expressions to our previous work\cite{Kim2021} for the local harmonic bath model. The energy conservation and detailed balance for these dissipation rate formulae [Eqs.~(\ref{eq:diss_eqn_loc})--(\ref{eq:FA_t})] are now guaranteed by the general proof presented in \stn{subsection:proof}, which was not attempted in the previous work\cite{Kim2021} due to the complicated form of the expressions.

The success of MQME-D implies that it is more fundamental and useful compared to the former approach based on explicit quantization of the bath mode. The approach enables establishing dissipation pathways without following explicitly the dynamics of the bath. At a formal level the projection operator technique enables us to develop practical expression without the need to overcome the technical challenge of adding up over all contributions due to individual bath components.\cite{Kim2021} We expect that the procedure illustrated herein to be applicable to a broad range of models for open quantum system dynamics, provided that we can concretely specify the form of individual bath components and their coupling to the subsystem.

We note, however, that the current version of MQME-D only focuses on the projected density $\hmc{P}\hat{\rho}$ and therefore any contributions from $\hmc{Q}\hat{\rho}$ and its interaction with $\hmc{P}\hat{\rho}$ are neglected. Moreover, the derivations were based on the second-order perturbation theory and the Markov approximation. Overall, these aspects would limit the accuracy of the theory under the presence of strong subsystem-bath interaction or underdamped bath modes that resonate with the subsystem. Hence, it would be desirable to continue the search for a more general theory of dissipation that can take the dynamics of $\hmc{Q}\hat{\rho}$ into account,\cite{Trushechkin2019} incorporate non-equilibrium relaxation of the bath,\cite{Jang2002,Jang2004} or target generalized quantum master equations\cite{Shi2003,Shi2004,Kelly2016,Amati2022} that do not involve any Markovianity or time-locality. Increasing the order of the perturbative treatment or even calculation of entropy production may be also attempted based on the previous studies along this direction.\cite{Laird1991,Reichman1996,Landi2021} We note that, when there are only a handful of bath components that exert strong subsystem-bath coupling or non-Markovianity, they can be directly dealt by forming the extended subsystem.\cite{Strasberg2018}

Another shortcoming of the current work is that it can only handle the situations where the bath exclusively couples to the subsystem Hamiltonian through the diagonal part. Nevertheless, we believe it can be extended in the near future toward more general situations involving both diagonal and off-diagonal couplings.

We expect that the power of MQME-D and its possible extensions will be demonstrated by combining our theory of dissipation with elaborate model Hamiltonians for molecular systems. For example, detailed spectral densities were constructed for natural and artificial molecular systems by fitting the linearly coupled harmonic bath model to spectroscopic signals\cite{Novoderezhkin2004,Ratsep2007,Gustin2023} or quantum chemistry-based calculations.\cite{Lee2017,Morrison2017,Kim2018,Lei2020} Our framework can be used to locate parts of these spectral densities that contribute the most to the dynamics, which may then be connected to specific features of the molecular vibration to yield comprehensive interpretations about how vibronic interaction affects non-adiabatic dynamics.

%Future prospects include ~ going beyond second order ~ entropy production~

%Our theory of dissipation allows us to draw a clear picture about microscopic factors involved in the open quantum system dynamics. In particular, the dissipative potential $\mc{I}_{BA}(\omega)$ enables us to concretely determine the presence of vibronic resonance, which would shed light on the discussions regarding the role of molecular vibrations in quantum transport. Our work implies that, by revisiting existing master equations and investigating their detailed structures, we may uncover valuable theoretical tools toward studying fundamental aspects of the quantum dynamics. Despite being often approximate, we envisage that such methods will be useful in disentangling the dynamics of complicated molecular systems which is out of the reach of numerically exact simulation methods.

%\section*{Supplementary Material}
%See the supplementary material for 

\begin{acknowledgments}
CWK was financially supported by Chonnam National University (Grant Number: 2022-2620) and the National Research Foundation of Korea (NRF) grants funded by the Ministry of Science and ICT (MSIT) of Korea (Grant Number: 2022R1F1A1074027, 2023M3K5A1094813, and RS-2023-00218219). IF is supported by the National Science Foundation under Grant No. CHE-2102386 and PHY-2310657. The authors thank Ignacio Gustin (University of Rochester) for his careful feedback on an earlier version of this manuscript. 
\end{acknowledgments}

\section*{Data Availability}
Data sharing is not applicable to this paper as no new data were created or analyzed in this study.

\appendix
\section{Derivation of \eq{eq:equivalence}}\label{section:equivalence}
In this appendix, we prove \eq{eq:equivalence} which forms the theoretical foundation for calculating the dissipation with the projection operator technique. We first establish a general identity which can treat a broader range of master equations, and then show that \eq{eq:equivalence} is a specific case of this identity.

Let us start from a general Hamiltonian for open quantum system
\begin{equation}\label{eq:H_global_gen}
    \hat{H}' = \hat{H}_\ti{sub} + \hat{H}_\ti{b1} + \hat{H}_\ti{b2},
\end{equation}
where $\hat{H}_\ti{sub}$ is the subsystem Hamiltonian given by \eq{eq:H_sys}. The remaining terms $\hat{H}_\ti{b1} + \hat{H}_\ti{b2}$ are the combined Hamiltonian for both the bath and the subsystem-bath interaction, where $\hat{H}_\ti{b1}$ describes the bath component of interest and $\hat{H}_\ti{b2}$ the rest. From now on, for the sake of mathematical rigor, we will explicitly express all operators for the system in the direct product form. The subsystem Hamiltonian accordingly becomes $\hat{H}_\ti{sub} = \hat{H}_\ti{sub} \otimes \hat{I}_\ti{b1} \otimes \hat{I}_\ti{b2}$ where
\begin{equation}\label{eq:H_sys_gen}
    \hat{H}_\ti{sub} = \sum_A E_A \ket{A}\bra{A} + \sum_{A,B} V_{AB} \ket{A}\bra{B},
\end{equation}
and $\hat{I}_\ti{b1}$ and $\hat{I}_\ti{b2}$ are the identity operators within the subspaces spanned by the first and second bath components, respectively. The general expressions for the two bath Hamiltonians $\hat{H}_\ti{b1}$ and $\hat{H}_\ti{b2}$ are
\begin{subequations}\label{eq:bc}
\begin{equation}
    \hat{H}_\ti{b1} = \hat{I}_\ti{sub} \otimes \hat{b}_\ti{b1} \otimes \hat{I}_\ti{b2} + \sum_{A,B} \big[ \ket{A} \bra{B} \otimes \hat{c}_\ti{b1}^{AB} \otimes \hat{I}_\ti{b2} \big],
\end{equation}
\begin{equation}
    \hat{H}_\ti{b2} = \hat{I}_\ti{sub} \otimes \hat{I}_\ti{b1} \otimes \hat{b}_\ti{b2} + \sum_{A,B} \big[ \ket{A} \bra{B} \otimes \hat{I}_\ti{b1} \otimes \hat{c}_\ti{b2}^{AB} \big],
\end{equation}
\end{subequations}
where $\hat{I}_\ti{sub}$ is the identity operator for the subspace spanned by the subsystem, and $\hat{b}_\ti{b1}$ and $\{ \hat{c}_\ti{b1}^{AB} \}$ ($\hat{b}_\ti{b2}$ and $\{ \hat{c}_\ti{b2}^{AB} \}$) are arbitrary operators that act on the subspace spanned by the first (second) bath component. We now conceive a general projection super-operator $\hmc{P}'$ which acts on the full density matrix of the system according to
\begin{equation}\label{eq:proj_dm_gen}
    \begin{split}
    \hmc{P}' \hat{\rho}'(t) = \sum_{A, B} \big[ \sigma_{AB}(t) \ket{A}\bra{B} \otimes \hat{R}_\ti{b1}^{AB} \otimes \hat{R}_\ti{b2}^{AB} \big].
    \end{split}
\end{equation}
In the above, we have introduced 
\begin{equation}
    \sigma_{AB}(t) = \bra{A} \tr_\ti{b1} \big[ \tr_\ti{b2} [\hat{\rho}'(t)] \big] \ket{B}
\end{equation}
as an element of the subsystem RDM, with $\hat{R}_\ti{b1}^{AB}$ and $\hat{R}_\ti{b2}^{AB}$ being the corresponding reference bath densities. As usual, the traces $\tr_\ti{b1}$ and $\tr_\ti{b2}$ indicate the trace over the subspaces spanned by the first and second bath components, respectively.

We now examine the condition required for
\begin{equation}\label{eq:equivalence_gen}
    \tr \bigg[ \hat{H}_\ti{b1} \frac{d \hat{\rho}' (t)}{dt} \bigg] = \tr \bigg[ \hat{H}_\ti{b1} \frac{d}{dt} [\hmc{P}_\ti{b2} \hat{\rho}' (t)] \bigg]
\end{equation}
to be valid, in which $\hmc{P}_\ti{b2}$ is the projection super-operator that only acts on the second bath component
\begin{equation}\label{eq:proj_b2}
    \hmc{P}_\ti{b2} \hat{O} = \sum_{A, B} 
    \bigg( \ket{A}\bra{B} \otimes \tr_\ti{b2} \big[ \bra{A} \hat{O} \ket{B} \big] \otimes \hat{R}_\ti{b2}^{AB} \bigg).
\end{equation}
Here, $\hat{O}$ is an arbitrary operator in the entire space spanned by $\hat{H}'$. If \eq{eq:equivalence_gen} is satisfied, it means that the rate of dissipation into a bath component is unaffected even when all other bath components are constrained by $\hmc{P}_\ti{b2}$. In \stn{subsection:calc_diss}, we exploited this property to bring out a bath component from the projection and calculate the dissipation associated with it, while still keeping other bath modes under the projection. To validate \eq{eq:equivalence_gen}, we first switch the order of the time-derivative and $\hmc{P}_\ti{b2}$ on the right-hand side, and then re-express the time-derivatives in each side by employing Liouville-von Neumann equation. The result is
\begin{subequations}\label{eq:equivalence_gen2}
\begin{equation}
    \tr \bigg[ \hat{H}_\ti{b1} \frac{d \hat{\rho}' (t)}{dt} \bigg] = i\hbar \: \tr \big[ \hat{H}_\ti{b1} \big\{ \hat{H}' \hat{\rho}'(t) - \hat{\rho}' (t) \hat{H}' \big\} \big],
\end{equation}
\begin{equation}
    \tr \bigg[ \hat{H}_\ti{b1} \frac{d}{dt} [\hmc{P}_\ti{b2} \hat{\rho}' (t)] \bigg] = i \hbar \: \tr \big[ \hat{H}_\ti{b1} \hmc{P}_\ti{b2} \big\{ \hat{H}' \hat{\rho}' (t) - \hat{\rho}' (t) \hat{H}' \big\} \big].
\end{equation}
\end{subequations}
where we have taken into account that there is no time-dependence in $\hmc{P}_\ti{b2}$. We now insert the explicit expressions for the Hamiltonian operators [Eqs.~(\ref{eq:H_global_gen})--(\ref{eq:proj_dm_gen})] in the right-hand side of Eq.~(\ref{eq:equivalence_gen2}a) and simplify the resulting expression to obtain
\begin{equation}\label{eq:lhs}
    \begin{split}
    \tr \bigg[ \hat{H}_\ti{b1} \frac{d \hat{\rho}' (t)}{dt} \bigg] &= \sum_A \tr_\ti{b1} \bigg[ \tr_\ti{b2} \big[ (\hat{b}_\ti{b1} \otimes \hat{I}_\ti{b2}) \bra{A} \hat{S}'(t) \ket{A} \big] \bigg] \\
    &+ \sum_{A,B} \tr_\ti{b1} \bigg[ \tr_\ti{b2} \big[ (\hat{c}_\ti{b1}^{AB} \otimes \hat{I}_\ti{b2}) \bra{B} \hat{S}'(t) \ket{A} \big] \bigg],
    \end{split}
\end{equation}
where we have used $\hat{S}'(t)$ to abbreviate $i \hbar \: [\hat{H}' \hat{\rho}' (t) - \hat{\rho}' (t) \hat{H}']$. A similar procedure can be also carried out for the right-hand side of Eq.~(\ref{eq:equivalence_gen2}b) with the definition of $\hmc{P}_\ti{b2}$ [\eq{eq:proj_b2}], yielding
\begin{equation}\label{eq:rhs}
    \begin{split}
    \tr \bigg[ \hat{H}_\ti{b1} \frac{d}{dt} [\hmc{P}_\ti{b2} \hat{\rho}' (t)] \bigg] &=\sum_A \tr_\ti{b1} \bigg[ \hat{b}_\ti{b1} \tr_\ti{b2} \big[ \bra{A} \hat{S}'(t) \ket{A} \big] \bigg] \\
    &+ \sum_{A,B} \tr_\ti{b1} \bigg[ \hat{c}_\ti{b1}^{AB} \tr_\ti{b2} \big[ \bra{B} \hat{S}'(t) \ket{A} \big] \bigg].
    \end{split}
\end{equation}
If we compare the right-hand sides of \eqs{eq:lhs}{eq:rhs}, their only difference is whether or not the bath operators $\hat{b}_\ti{b1}$ and $\hat{c}_\ti{b1}^{AB}$ are within the trace over the subspace of the second bath component. A sufficient condition for taking the bath operators in \eq{eq:lhs} out of the trace is that $\bra{A}\hat{S}'(t)\ket{B}$ must be factorizable into operators within the individual bath subspaces. As every term in the system Hamiltonian [\eqs{eq:H_global_gen}{eq:bc}] are already in the direct product form, the only remaining requirement is that the projected densities $\bra{A} \hat{\rho}'(t) \ket{B}$ are factorizable. In general, such a condition is not satisfied because the subsystem-bath interaction [\eq{eq:bc}] entangles different bath components as the system evolves in time. Nevertheless, this issue can be remedied by making an additional assumption that $\hat{\rho}'(t) = \hmc{P}_\ti{b2} \hat{\rho}'(t)$ [\eq{eq:proj_b2}] at this instance. As this condition must be satisfied with classifying every single bath component as $\hat{H}_\ti{b1}$, it is required that $\hat{\rho}'(t) = \hmc{P}' \hat{\rho}'(t)$, which will remain valid at all instance if we keep our focus on the projected density (\stn{subsection:calc_diss}).

Having confirmed the validity of \eq{eq:equivalence_gen} within the formulation presented in our work, we can now observe that \eq{eq:equivalence} is a specific case of \eq{eq:equivalence_gen}, as $\hmc{P}'$ [\eq{eq:proj_dm_gen}] reduces to $\hmc{P}$ [\eq{eq:P_def}] when $\hat{R}_\ti{b1}^{AB} = \hat{r}_{Aj} \delta_{AB}$ and $\hat{R}_\ti{b2}^{AB} = \hat{R}_{A, j-} \delta_{AB}$. The connections between \eq{eq:H_global_gen} and the Hamiltonian models in \stn{section:specific} are also readily established. For example, the Hamiltonian for linearly coupled harmonic oscillator bath [\eqs{eq:H_bath_ho}{eq:H_int_ho}] is obtained by setting
\begin{subequations}\label{eq:bc_special}
\begin{equation}
    \hat{b}_\ti{b1} = \frac{\hat{p}_j^2}{2} + \frac{\omega_j^2}{2} \hat{x}_j^2, \quad \hat{c}_\ti{b1}^{AB} = \delta_{AB} \bigg( -\omega_j^2 d_{Aj}\hat{x}_j + \frac{\omega_j^2 d_{Aj}^2}{2} \bigg),
\end{equation}
\begin{equation}
    \hat{b}_\ti{b2} = \sum_{k \neq j} \bigg( \frac{\hat{p}_k^2}{2} + \frac{\omega_k^2}{2} \hat{x}_k^2 \bigg), \quad \hat{c}_\ti{b2}^{AB} = \delta_{AB} \sum_{k \neq j} \bigg( -\omega_k^2 d_{Ak}\hat{x}_k + \frac{\omega_k^2 d_{Ak}^2}{2} \bigg).
\end{equation}
\end{subequations}

\section{Validation of \eq{eq:equiv_dynamics}}\label{section:proof_equiv}
This Appendix presents the proof of \eq{eq:equiv_dynamics} which assures that the dynamics of the state populations is unaltered by calculation of the dissipation under the split of the projection operator $\hmc{P} = \hat{p}_j \hmc{P}_{j-}$ [\eq{eq:proj_factorize}]. We start by applying $\hat{p}_j$ from the left of \eq{eq:dproj_m_rho_dt},
\begin{equation}\label{eq:P_minus_expression}
    \hat{p}_j \frac{d}{dt} [\hmc{P}_{j-} \hat{\rho}(t)] = -\frac{1}{\hbar^2} \int_0^\infty \hat{p}_j \hmc{P}_{j-} \hmc{L}_1 \exp(-i t' \hmc{L}_0 / \hbar) \hmc{L}_1 \hmc{P}_{j-} \hat{\rho}(t) \: dt',
\end{equation}
which is justified by that $\hat{p}_j$ is time-independent and therefore the order of the projection and integration can be switched. Then, the integrand of \eq{eq:P_minus_expression} can be simplified by using \eq{eq:integrand} and explicitly applying $\hat{p}_j$ based on Eq.~(\ref{eq:proj_factorize}a),
\begin{equation}\label{eq:integrand2}
    \begin{split}
    &\hat{p}_j \hmc{P}_{j-} \hmc{L}_1 \exp(-i t' \hmc{L}_0 / \hbar) \hmc{L}_1 \hmc{P}_{j-} \hat{\rho}(t) = \sum_A \sum_{B \neq A} \bigg[ |V_{AB}|^2 \ket{A} \bra{A} \\
    &\otimes \bigg( P_A(t) \tr_{\ti{b}} [ \hat{U}_B \hat{R}_A \hat{U}_A^\dagger ] - P_B(t) \tr_{\ti{b}} [ \hat{U}_B \hat{R}_B \hat{U}_A^\dagger 
    ] \bigg) \otimes \hat{R}_A \bigg] + \ti{H.c.}
    \end{split}
\end{equation}
We now insert this result in \eq{eq:P_minus_expression} and use \eq{eq:FRET_rateconst} to relate the integrals to the rate constants. The result is
\begin{equation}\label{eq:equiv_last}
    \begin{split}
    \hat{p}_j \frac{d}{dt} [\hmc{P}_{j-} \hat{\rho}(t)] &= \sum_A \sum_{B \neq A} \bigg([-K_{BA} P_A(t) + K_{AB} P_B(t)] \ket{A}\bra{A} \otimes \hat{R}_A\bigg) \\
    &= \sum_A \dot{P}_A(t) \ket{A}\bra{A} \otimes \hat{R}_A,
    \end{split}
\end{equation}
where the last equality follows from \eq{eq:FRET_rateeqn}. \Eq{eq:equiv_dynamics} can now be proven by connecting \eq{eq:equiv_last} to the time derivative of \eq{eq:P_def}.

\providecommand{\noopsort}[1]{}\providecommand{\singleletter}[1]{#1}%

\end{document}